\input amstex.tex
\magnification\magstep1
\documentstyle{amsppt}

%\redefine\Bbb{\bold}

\define\a{\alpha}

\define\g{\gamma}
\define\G{\Gamma}
\redefine\o{\omega}
\redefine\O{\Omega}
\redefine\l{\lambda}

\define\gm{\bold g}

\define\<#1,#2>{\langle #1,#2\rangle}
\define\dep(#1,#2){\text{det}_{#1}#2}
\define\norm(#1,#2){\parallel #1\parallel_{#2}}

\define\ep{\epsilon}
\define\psl{p\!\!\!\slash}
\define\Res{\text{Res}}
\topmatter
\title Vacuum polarization and the geometric phase: gauge invariance
\endtitle
\author Jouko Mickelsson \endauthor
\affil Department of Theoretical Physics, Royal Institute of Technology,
S-10044 Stockholm, Sweden \endaffil
\endtopmatter
\NoBlackBoxes

\document

ABSTRACT: A nonperturbative approach to the vacuum polarization for
quantized fermions in external vector potentials is discussed.
It is shown that by a suitable choice of counterterms the vacuum
polarization phase is both gauge and renormalization independent,
within a large class of nonperturbative renormalizations.

\vskip 0.4in

1. INTRODUCTION \newline\newline
For Dirac fermions in (non second quantized) external Yang-Mills fields
the 1-particle
scattering matrix $S$ is a well-defined unitary operator and moreover it has
a canonical second quantization $\hat S$ operating in the fermionic Fock
space [1]; for twice differentiable vector potentials one only
needs to assume certain fall-off properties at spatial and time infinity.

On the other hand, if one tries to expand the quantum scattering matrix
in ordinary Dyson-Feynman perturbation series one meets the well-known
vacuum polarization divergences, which must be taken care by suitable
(infinite) subtractions. Actually, it is only the phase of the vacuum
expectation value $<0|\hat S|0>$ which is diverging; the absolute value
is uniquely defined and finite by canonical quantization.  The crucial
point is that when passing from $S$ to $\hat S$ using the rules of canonical
hamiltonian quantization the phase remains ill-defined. It is exactly this
quantity which is diverging in perturbation theory.

One of the basic principles of relativistic quantum field theory is locality.
The fields are supposed to either commute (bosons) or anticommute (fermions)
at space-like separations. This is the case at least in four or higher
space-time dimensions; in lower dimensions there are more alternatives.
For this reason one prefers field theories in which the action is a local
differential polynomial of the fields. On the other hand, the requirement
that the action is a function of the fields and their derivatives up to
a finite degree at the same space-time point is by no means a necessary
condition for the locality in the above sense. In this paper we continue
the work on a nonperturbative renormalization which breaks the locality
in the narrow sense (the renormalized action contains field derivatives
up to infinite degree) but nevertheless is local in the framework of
Wightman axioms.

This work was initiated in [2] and further generalized in
[3]. In the previous paper [3] the question of gauge and renormalization
independence was left open. We fill the gap in the present paper in the case
of infinitesimal gauge and renormalization variations.
\vskip 0.3in

2. THE RENORMALIZATION \newline\newline

We shall study massless Dirac fermions coupled to a gauge potential $A$
in Minkowski space. The potential is a smooth 1-form $A_{\mu}(x)\, dx^{\mu}$ in
space-time with
values in the Lie algebra $\gm$ of a compact gauge group $G.$ The elements
of $\gm$ are represented by hermitean (according to physics literature
convention) matrices in the complex vector space $\Bbb C^N.$ The free Dirac
operator is then $i\g^{\mu} \partial_{\mu}.$ The metric is $x^{\mu} x_{\mu}=
x_0^2 -x_1^2 - \dots -x_d^2.$ The Dirac gamma matrices satisfy
$\g_{\mu}\g_{\nu}
+\g_{\nu}\g_{\mu} = 2 g_{\mu\nu},$ $\g_0$ is hermitean and $\g_k$ is
antihermitean for $k\neq 0.$ We fix a hermitean matrix $\G$ such that $\G^2=1$
and $\G\g_{\mu}
=-\g_{\mu}\G.$ The chiral projectors are $P_{\pm} = \frac12(\G \pm 1).$

The Dirac hamiltonian in background gauge field $A$ is
$$D_A = -\g^0\g^k(i\partial_k +A_k) -A_0.$$
We shall assume that $A(x)$ and its derivatives vanish faster than
$|x|^{-d/2}$ when $|x| \to \infty.$

The one-particle scattering operator $S$ is defined as the limit of the
time evolution operator in the interaction picture, $U_I(t,-t) \to S$ as
$t\to\infty.$ The time evolution in the Schr\"odinger picture is defined by
$$i\partial_t U(t,t_0) = h(t) U(t,t_0) \text{ with } U(t_0,t_0)=1, \tag1$$
and in the interaction picture by
$$i\partial_t U_I(t,t_0) = V_I(t) U_I(t,t_0), \text{ with } U_I(t_0,t_0)=1.
\tag2$$
The interaction is  $V_I(t) = e^{it h_0} V(t) e^{-it h_0},$ the total
hamiltonian being $h(t)= D_A= h_0 +V(t)$ with $h_0 =D_0 = -i\g_0\g^k \partial_k.$
The quantum divergences are related to the
fact that when $V= -\g^0\g^k A_k(\bold x,t) -A_0(\bold x,t)$ is the interaction
with a Yang-Mills potential then the quantization of $\hat U_I(t,-\infty)$ for
intermediate times $t < \infty$ is not well-defined. Let $\ep=h_0/|h_0|$ be the
sign of the free hamiltonian. In general, a 1-particle
operator $K$ has a canonical quantization in the Fock space of free fermions iff
$[\ep, K]$ is Hilbert-Schmidt. For a review of this and other properties of
representations of CAR algebra, see [9].
Generically, the Hilbert-Schmidt property is not satisfied when $K=V(t).$

Let us recall the basic facts about (free) Fock space quantization of the canonical
anticommutation relations. To each vector $u\in H$ in the 1-particle Hilbert
space one associates a creation operator $a^*(u)$ and an annihilation operator
$a(u),$ the latter depending antilinearly on the parameter and the former
linearly, with the only nonzero anticommutation relations
$$a^*(u) a(v) + a(v) a^*(u)= <v,u>$$
where $<v,u>$ is the inner product in $H.$ Let $H=H_+\oplus H_-$ be splitting to
a pair of closed infinite-dimensional subspaces. In our discussion below $H_+$
will be the spectral subspace corresponding to nonnegative energies with respect
to the free Dirac hamiltonian $D_0.$ There is a unique (up to
equivalence) irreducible representation of the CAR algebra in a Hilbert space
$\Cal F$ such that
$$a^*(u)|0> =0 = a(v) |0> \text{ for $u\in H_-$ and $v\in H_+$ } $$
where $|0> \in \Cal F$ is a normalized vector, the vacuum.

Let $a^*_i, a_j,$ with $i,j\in\Bbb Z,$ be a complete set of
creation and annihilation operators such that the index $i\geq 0$ corresponds
to nonnegative energies and $i<0$ to negative energies,
$$a^*_i a_j +a_j a^*_i = \delta_{ij}.$$
The normal ordering is defined by $:a^*_i a_j : = a^*_i a_j$ except when $i=j
<0$ and then $:a^*_i a_j: = - a_j a^*_i.$
If $X,Y$ is a pair of one-particle
operators then the canonical quantizations
$$\hat X = \sum X_{ij}  : a^*_i a_j : $$
satisfy the algebra
$$[\hat X,\hat Y] = \widehat{[X,Y]} +\o(X,Y),$$
and the quantum operators are defined such that $[\hat X, a^*_i] =\sum X_{ji}
a^*_j .$ The 2-cocycle $\o$ is defined by, [10],
$$\o(X,Y) = \frac14 \text{tr}\,\ep [\ep,X][\ep,Y].\tag3$$

Here we meet again the Hilbert-Schmidt condition: The 2-cocycle is defined
only for bounded operators $X$ such that $[\ep,X]$ is Hilbert-Schmidt.

The starting point for our discussion here is the following renormalization
which makes $U_I$ quantizable, [2,3]. For each (smooth) potential $A=A_0 dt
+A_k dx^k$ one
defines a unitary operator $T_A$ in the 1-particle space with the following
property.
Define $U_{ren}(t,t_0) = T_A(t) U_I(t,t_0) T_A(t_0)^{-1}.$ Then 1)
$[\epsilon, U_{ren}(t,t_0)]$ is Hilbert-Schmidt, 2) $T_A(t) \to 1$ as $t\to
\pm \infty.$  The last property quarantees that the renormalization does not
affect the scattering matrix $S$ whereas the first condition guarantees that
the renormalized time evolution is quantizable in the free Fock space.

The Hilbert-Schmidt condition on operators $[\epsilon,g]$ defines the
restricted unitary group $U_1$ of unitary operators $g,$ [8]. The second
quantization of elements in $U_1$ defines a central extension $\hat U_1$ as
discussed in detail in [8]. Now we have a smooth path of operators $g(t)=
U_{ren}(t)=
U_{ren}(t,-\infty)$ in $U_1$ with the initial condition $g(-\infty)=1$
and $g(+\infty)=S.$ The central extension of $U_1$ defines a natural
connection in the circle bundle $\hat U_1 \to U_1.$ The curvature of this
connection has a simple formula: It is simply given as $curv(X,Y)=\o(X,Y)$
where $X,Y$ are tangent vectors at $g\in U_1,$ identified through left
translation on the group as elements of the
Lie algebra of $U_1.$ The phase of $\hat S$ is defined \it through parallel
transport, \rm with respect to the connection above.

Our definition of phase of $\hat S$ is \it causal: \rm A scattering process
in an external field $A$ followed by the scattering in $A'$ defines the same
phase as the scattering for $A''=A\cup A',$ when both $A,A'$ have finite
nonoverlapping support
in time and $A''$ denotes a field $A''(t)=A(t)$ for $t <t_0$ and $A''(t)=
A'(t)$ for $t>t_0;$ here $t_0$ separates the supports of $A,A'.$
                               \define\Asl{A\!\!\!\!\slash}
The operator $T_A$ is not uniquely defined. It is more convenient to define
the transformation first in the Schr\"odinger picture (1).
A simple formula which works  is (here $\Asl= \g^0\g^k A_k$ and $E\!\!\!\!\slash
=\partial_t \Asl-\psl A_0 +[A_0,\Asl]$)
$$T_A = \exp\left(\frac14 [\frac{1}{D_0}, \Asl] -
\frac18[\frac{1}{D_0}\Asl\frac{1}{D_0},\Asl ]-\frac{i}{4} \frac{1}{D_0}
E\!\!\!\!\slash \frac{1}{D_0} \right), \tag4  $$
where it is understood that the singularity at the zero modes of $D_0$ is taken
care of by an infrared regularization, for example $\frac{1}{D_0} \to
\frac{D_0}{D_0^2 + \a^2}$ for some nonzero real number $\a.$ In the interaction
picture one uses the operator $\exp(ith_0) T_A \exp(-ith_0).$
Note that this choice commutes with the chiral projection operators $P_{\pm}.$

For the proof of validity of the choice (4) it is convenient to use the symbol
calculus for pseudodifferential operators.  A PSDO $B$ acting on vector valued
functions in $\Bbb R^d$ is represented by a smooth matrix valued function
$b(x,p)$ of coordinates
$x_i$ and momenta $p_i.$ The operators which we need have an asymptotic
expansion
$$b(x,p) \sim b_k + b_{k-1} + b_{k-2} +\dots$$
where each $b_j$ is a smooth function of $x,p$ for $|p| >1,$ homogeneous of
degree $j$ in the momenta. The asymptotic expansion   for the product of two
operators $A,B$ is obtained from the formula
$$a*b = \sum \frac{(-i)^{|m|}}{m!} \frac{\partial^m a}{\partial x^m}
\frac{\partial^m b}{\partial p^m}$$
where the multi-index notation $m=(m_1,m_2,\dots,m_d)$ is used; $|m| =m_1+m_2
+\dots m_d,$ $m! = m_1! m_2!\dots m_d!$ etc. In particular, the degree of the
highest term in the product is the sum of degrees of the factors. Each
differentiation in momentum space lowers the degree of the operator by one.

If $A$ is an operator with symbol $a$ such that $a(x,p) \to 0$ faster than
$|x|^{-d}$ as $|x| \to \infty$ and deg$\,a < -d$ then $A$ is a trace-class
operator and
$$\text{tr}\, A = \frac{1}{(2\pi)^d} \int \text{tr}\, a(x,p) dx dp.$$
We shall now assume that the components $A_{\mu}$ of the vector potential
satisfy the above boundary condition at $|x| \to \infty.$

The time evolution equation
for $U_{ren}(t,t_0) = T_A(t) U(t,t_0)T_A(t_0)^{-1}$ in the Schr\"odinger
picture is
$$i\partial_t U_{ren}(t,t_0) = (h_0 +W(t)) U_{ren}(t,t_0)$$
with
$$W(t) = (i\partial_t T_A) T_A^{-1}  +T_A (h+V(t)) T_A^{-1}.\tag5$$
Expanding the exponential and arranging terms according to powers of the
inverse of momentum (i.e., of $D_0$,) one gets $[\ep, W]= R_1 +R_2 +\dots,$ where
the dots denote terms which behave explicitly as $|D_0|^k$ with $k\leq -2$
for high momenta, and                         \define\DD{\frac{D_0}{|D_0|}}
  $$R_1= \frac12\frac{D_0}{|D_0|}\Asl -\frac12 \Asl \DD +\frac14 |D_0|\Asl\frac{1}
{D_0} -\frac14 D_0 \Asl\frac{1}{|D_0|} +\frac14 \frac{1}{|D_0|} \Asl D_0
-\frac14 \frac{1}{D_0}\Asl |D_0|       $$
and the second term, which is quadratic in $\Asl$, is \define\DI{\frac{1}{D_0}}
$$\align R_2 =&\frac14 \DI [|D_0|,\Asl^2] \DI +\frac14 \left[\frac{1}{|D_0|},
\Asl^2\right]
+\frac18 D_0 \left[\Asl\DI\Asl,\frac{1}{|D_0|}\right] +\\ &+\frac18
\left[\Asl\DI\Asl, \frac{1}{|D_0|}\right] D_0
+\frac18 \DI \left[\Asl\DI\Asl, |D_0|\right] +\frac18 \left[\DI\Asl\DI,
|D_0|\right] \DI.\endalign$$
Since the commutator $[|D_0|^k,\Asl]$ is of order $|D_0|^{k-1}$ in momenta,
all terms in $R_2$ are actually of order $-2$ or less. In order to estimate
$R_1$ we write it in equivalent form
$$R_1 = \frac14 \DD \left[ [\Asl, |D_0|], \frac{1}{|D_0|} \right]
-\frac14 \left[ [\Asl, |D_0|], \frac{1}{|D_0|}\right] \DD$$
and observe both terms are of order $-2$ for the same reason as in the case
of $R_2.$ We have disregarded all the low order terms because in three space
dimensions  the condition that a PSDO is Hilbert-Schmidt (which was
required for canonical quantization) is precisely the requirement that the
operator vanishes for high momenta faster than $|p|=|D_0|$ raised to power
$-3/2.$ Thus after our renormalization (= conjugation by the time-dependent
unitary operator $T_A$) the gauge interaction can be lifted to a finite
operator in the fermionic Fock space.

The total renormalized hamiltonian is now a sum of the free (unbounded)
self-adjoint hamiltonian and the bounded self-adjoint interaction. By the
Kato-Rellich theorem [11] the total hamiltonian is self-adjoint and according to
Stone's theorem it defines the time evolution  as a strongly
continuous unitary one-parameter group in the Fock space.

This method can be extended to other interactions as well, under very
general conditions [3]. It has been also used to derive the chiral anomaly
in the hamiltonian framework [2], [3].

The curvature formula  (3) is   \it nonlocal. \rm The computation of the
trace involves space derivatives up to arbitrary high power because of the
Green's function $1/D_0$ in the renormalization prescription. However, the
curvature is equivalent in cohomology to a local formula. We use the residue
formula, [7],
$$\Res\, B = \frac{1}{(2\pi)^d} \int_{|p|=1} \text{tr}\, b_{-d}(x,p)\, d^d x
d^{d-1} p.$$
It behaves like trace, $\Res[A,B] =0.$ Let $\gm$ be the Lie algebra of bounded
pseudodifferential operators satisfying the conditions 1) that the degree of the
commutator $[\ep,X]$ is strictly less than $-d/2$ (here the space dimension $d  =3$),
2) the symbol of $X$ and its derivatives decrease faster than $|x|^{-d/2}$
as $|x| \to\infty.$  We
have  then
\proclaim{Theorem 1} Define the following 2-cocycle on $\gm:$
$$\o_{loc}(X,Y)=  \text{Res}\,\ep [\text{log} |p|,X] Y,\tag6$$
where $|p|,$ the length of three momentum, is the symbol of $|D_0|.$
Then the difference $\o -\o_{loc}$ is a coboundary. \endproclaim
For a proof see [4] (for the special case of renormalized current operators
[2]). Actually, in [4] it was assumed that the manifold is compact. This
was needed for the convergence of the integral over $x$ in the trace formula
for PSDO's. Our boundary condition does the same job.

Without the sign $\ep$ this formula would give the Radul cocycle, which is a
2-cocycle on the algebra of \it all \rm PSDO's on a compact manifold [5]. In
quantum field theory it is important to keep the sign because this is the
cocycle arising from normal ordering.
The equivalence of the two cocycles means simply that if $\tilde X=
\hat X +\l(X)$ for a suitable complex linear form on the algebra of one-particle
operators then
$$[\tilde X, \tilde Y] = \widetilde{[X,Y]} + \o_{loc}(X,Y).\tag7$$

The great advantage of the residue formula is that it only depends on the term
in
the appropriate PSDO which is precisely of asymptotic degree $-d$ (here $-3$)
when expanded asymptotically in powers of $1/|p|.$ In operator products each
derivative in configuration space is associated with a derivative in momentum
space. Since a differentiation in momentum space decreases the homogeneous
degree of the operator it follows that a calculation of the residue can
involve derivatives with respect to $x_i$'s only up to a finite order.

\vskip 0.3in
3. GAUGE INDEPENDENCE \newline\newline

The prescription for the vacuum polarization phase in section 2 is not
gauge independent. Furthermore, if we change the renormalization operator
$T_A$ the phase will be changed, too. Of course, in the case of chiral
fermions the gauge dependence is to be expected because of the chiral
anomaly; this was discussed in [3].

In the case of Dirac fermions we expect on topological grounds that one
should be able to construct counterterms in the hamiltonian such that the
gauge variance of the counterterms exactly balances the gauge variation
of the phase, so that the total phase would be gauge invariant. We shall
confirm this explicitly below in the case of infinitesimal variations.
At the same time we shall also get rid of
the renormalization dependent part of the phase.

Suppose that we have constructed another renormalization operator $T'_A$ which
commutes with the chiral projection operators and with $\g_0,$ such
that  each term in the asymptotic expansion
$$T'_A = 1 + t_{-1}(A) + t_{-2}(A) +\dots,$$
in homogeneous terms $t_k$ of degree $k$ in momenta, is a local differential
polynomial in the components of the vector potential $A.$ The condition that
the renormalization commutes with $\g_0$ means that the action of $T'_A$
on left chiral sector is equivalent to the action on the right-hand spinors.

The change of
phase of the quantum scattering is given by a parallel transport around
a loop of time evolution operators obtained by first travelling the path
$g(t)$ from the point $g(-\infty)=1$ to $g(\infty)=S$ defined by the renormalized
time evolution due to $T_A$ and then following backwards in time the time
evolution defined by the renormalization $T'_A.$  But the logarithm of the
parallel transport phase is equal to the integral of the curvature over
a surface bounded by the loop.

In the following we shall use the local curvature formula (6). This has the
advantage that the parallel transport around a closed loop
is given by a space integral of a local differential
expression in the vector potential. This is in the spirit of local quantum
field theory: A change in the renormalization corresponds to a local counterterm
in the Lagrangian.

Since all our operators, the (renormalized) hamiltonian, the renormalization
operator, and gauge transformations, are composed of a pair of operators
$(K_L, K_R)$ acting on the two chiral sectors we can split the curvature
into chiral constituents as follows. The left and right spinors are
indentified through the unitary transformation $\g_0$ so that we can think of
$K_L$ and $K_R$ as acting in the same space. Thus we may define $K_{\pm} =
K_L \pm K_R.$ Since the sign $\ep$ is odd, $\ep_L = -\ep_R,$ we can write
$$2\o(X,Y) = \o_L(X_-,Y_+)+ \o_L(X_+,Y_-).$$
Here $\ell=\text{log}|p|$ is the symbol of $h_0$ and $\o_L(X;Y) =
\Res\, \ep_L [\ell,X_L] Y_L .$
We want to show that this is equivalent to a 2-form $\o_-$ such that
$\o_-(X,Y)=0$ if either $X_-=0$ or $Y_-=0.$ We denote $\Res'\, X = \Res\,
\ep_L X.$

Let $GL_1$ denote the group of all invertible bounded PSDO's $g$ in $H$
such that the degree of $[\epsilon,g]$  is strictly less than $-d/2.$ The
Lie algebra of $GL_1$ is then equal to $\gm.$ Let $GL_{1,c}$ be the subgroup
of $GL_1$ consisting of the
chirally diagonal operators $f=(f_L,f_R).$

The time evolution operator in Schr\"odinger picture does not have a good
asymptotic expansion in inverse powers of momenta and therefore it is not
immediately clear that one can define the Wodzicki residue for a product
involving $U(t)$ as a factor.
For example, the free time evolution $\exp(it\psl)$ contains arbitrary high
powers of momenta and therefore it is not a classical PSDO. However, the time
evolution
$U_I(t)=U_I(t,-\infty)$ in the interaction picture has a well behaved symbol.
The leading terms (in powers of momenta)
are obtained from the Dyson expansion in which the individual
terms are composed of products of the Green's function of the Dirac
operator and the interaction hamiltonian. This gives directly a classical
asymptotic expansion for the PSDO $U_I(t).$

In the following we deal with the 'Maurer-Cartan' form $\theta=[\ell,f]f^{-1}.$
This is a classical PSDO when $f= U_I(t).$ In the case of the Schr\"odinger
picture it is replaced by $\exp(ith_0) \theta \exp(-ith_0).$ The computation
of residues for this type of operators involves oscillatory integrals. These
can be handled by Guillemin's method for Fourier integral operators, [12].
For this reason the curvature formula extends to the bigger group $GL_F$
which is the smallest group containing $GL_1$ and the free time
evolution operators $\exp(ith_0).$

\proclaim{Lemma} The restriction of the curvature $\o_{loc}$ to the chirally
diagonal subgroup $GL_{F,c}\subset GL_F$
is equivalent in de Rham cohomology to the 2-form
$$\o'(X,Y)=\o_-(X_-,Y_-)= -\Res'[\ell,X_-] Y_-  +
\Res' [\ell,f_L]f_L^{-1} [X_-,Y_-]. $$
\endproclaim
\demo{Proof}  Define the 1-form
$$\phi(f;X)=  \Res'  [\ell,f_L]f_L^{-1} X_-$$
on the group $GL_{F,c}.$ The exterior derivative of this is computed through
$$(d\phi)(X,Y) = \Cal L_X \phi(f;Y) -\Cal L_Y\phi(f;X) -\phi(f;[X,Y])$$
using the definition of the left action, $\Cal L_X F(f)= \frac{\partial}{
\partial t} F(e^{-tX}f)|_{t=0}.$ The result is
$$(d\phi)(X,Y)= -\Res'[\ell,X_-] Y_-  -\Res' [\ell,X_L] Y_L
+\Res'[\ell,X_R] Y_R +\Res' [\ell,f_L]f_L^{-1} [X_-,Y_-].$$
Adding the right-hand-side to $\o(X,Y)$ gives $\o_-(X,Y).$
\enddemo

We denote by $q(R)$ the canonical second quantization of an operator $R$ and
$W=W(A,T)$ is the renormalized interaction (5).
\proclaim{Theorem 2} The modified quantum hamiltonian
$$q'(h_0 +W) = q(h_0 +W) + \phi(U(A;t); h_0 +W)$$
has the property that the phase of the quantum scattering matrix is
invariant under infinitesimal gauge transformations. Moreover, the phase
is invariant with respect to all infinitesimal variations $X$ of $T_A$ such
that 1) $X$ is a PSDO in the Lie algebra of $U_1,$ 2) $X$ is chirally even,
i.e. it commutes with $\g_0$ and $\G.$ \endproclaim
\demo{Proof} The second term on the right-hand-side  means that we compute the
quantum phase
using a modified connection, corresponding to adding to the old connection
(with curvature $\o_{loc}$) the 1-form $\phi.$
In general, a variation of the parallel transport phase defined by a connection
$\theta$ on any manifold is given by
$$ \Cal L_X \int \theta(\gamma'(t)) dt = \int \Omega_{g(t)} (\gamma'(t), X)
dt$$
where $\O$ is the curvature form and $X(t)$ is a vector field along the curve
$\gamma(t).$ Applying this to the case when $\gamma(t) = U(t),$
(the tangent vector $\gamma'(t)$ is replaced by the element $h_0 + W= iU'(t)
U(t)^{-1})$) and $X$ is an infinitesimal gauge transformation shows that at time
$t$ the infinitesimal gauge variation of the phase is
$$\Cal L_X (phase)=\omega'(h_0 +W, X).\tag8$$
But now $X_-=0$ and therefore this expression vanishes and thus also the total
gauge variation of the quantum phase as an integral with respect to $t$ of
(8).
\enddemo
\bf Remark \rm We started from the standard minimal coupling of fermions to
external gauge fields. We performed a renormalization by the field dependent
unitary operator $T_A.$  The new interaction hamiltonian, although nonlocal,
is still unitarily equivalent to the original local interaction.
There is a very natural generalization of our discussion in the spirit
of noncommutative geometry [6]. The original interaction $\gamma^{\mu} A_{\mu}$
could be replaced by any bounded pseudodifferential operator without
affecting our results (or even by a larger class of operators, see [3]).
This would lead to a hamiltonian counterpart of the noncommutative
Yang-Mills action functional. It would be an intermediate step between
the ordinary Yang-Mills theory and the abstract universal Yang-Mills theory
proposed by Rajeev, [13].

\vskip 0.2in
\bf References \rm \newline\newline
\noindent [1] J. Palmer: Scattering automorphisms of the Dirac field,
J. Math. Anal. Appl \bf 64, \rm 189-215 (1978). S.N.M Ruijsenaars: Charged
particles
in external fields; I, classical theory, J. Math Phys.
\bf 18, \rm 720-737 (1976). Gauge invariance and implementability of the S-operator
for spin-0 and spin-1/2 particles in time-dependent classical fields, J. Funct.
Anal. \bf 33, \rm 47-57 (1979)\newline\newline
[2] J. Mickelsson: Wodzicki residue and anomalies of current algebras,
\it Integrable Systems and Strings, \rm
ed. by A. Alekseev et al., Springer Lecture Notes in Physics \bf 436, \rm pp.
123-135 (1994).
Regularization of current algebra,
\it Constraint Theory and Quantization Methods, \rm ed. by F. Colomo, L. Lusanna, and
G. Marmo, World Scientific, pp. 72-79 (1994). Renormalization of current
algebra, \it Generalized Symmetries in Physics, \rm
ed. by H.-D. Doebner, V.K. Dobrev, and A.G. Ushveridze, World Scientific,
pp. 328-337 (1994)
\newline\newline
[3] E. Langmann and J. Mickelsson: Scattering matrix in external field problems,
J. Math. Phys. \bf 37, \rm 3933-3953 (1996)                  \newline\newline
[4] M. Cederwall, G. Ferretti, B.E.W. Nilsson, A. Westerberg: Schwinger terms
and cohomology of pseudodifferential operators,
Commun. Math.
Phys. \bf 175, \rm  203-220 (1996)       \newline\newline
[5] A.O. Radul: Lie algebra of differential operators, their central extensions,
and $\Cal W$-algberas, Funct. Anal. Appl. \bf 25, \rm  33-49 (1991)
\newline\newline
[6] A. Connes: \it Noncommutative geometry. \rm
Academic Press; London, San Diego (1994)   \newline\newline
[7] M. Wodzicki: Noncommutative residue,  \it
K-theory, arithmetic and geometry. \rm (ed. by Yu.I. Manin)
Lect. Notes in Math. \bf 1289, \rm pp.
320-399, Springer-Verlag (1987). Local invariants of spectral asymmetry,
Invent. Math. \bf 75, \rm 143-177 (1984).  \newline\newline
[8] A. Pressley and G. Segal: \it Loop Groups. \rm Clarendon Press, Oxford
(1986)\newline\newline
[9] H. Araki: Bogoliubov automorphisms and Fock representations of
canonical anticommutation relations. In: \it Contemporary Mathematics.
\rm  vol. \bf 62, \rm pp. 23-141, American Mathematical Society, Providence
(1987)\newline\newline
[10] L.-E. Lundberg: Quasi-free "second quantization",  Commun. Math. Phys.
\bf 50, \rm 103-112 (1976)\newline\newline
[11] M. Reed and B. Simon: \it Methods of Modern Mathematical Physics II, \rm
p. 162. Academic Press; London, San Diego (1975) \newline\newline
[12] V. Guillemin: Residue traces for certain algebras of Fourier integral
operators,  J. of Funct. Anal. \bf 115, \rm 391-417 (1993); Gauged
Lagrangian distributions, Advances in
Math. \bf 102, \rm 184-201 (1993)\newline\newline
[13] S.G. Rajeev: Universal gauge theory, Phys. Rev. \bf D42, \rm 2779-2791
(1990)

\enddocument